\documentclass[10pt,conference]{IEEEtran}
\IEEEoverridecommandlockouts
\usepackage{cite}
\usepackage[utf8]{inputenc}
\usepackage{amsmath,amssymb,amsfonts}
\usepackage{algorithmic}
\usepackage{graphicx}
\usepackage{textcomp}
\usepackage{xcolor}
\usepackage{listings}
\usepackage{makecell}
\usepackage{array}
\usepackage{booktabs}
\usepackage{soul}
\usepackage{svg}
\usepackage{cuted}
\usepackage{caption}
\usepackage{dblfloatfix}
\usepackage[most]{tcolorbox}

\lstdefinestyle{vscode}{
    backgroundcolor=\color{white},   
    commentstyle=\color{gray}\itshape,
    keywordstyle=\color{blue},
    stringstyle=\color{orange},
    identifierstyle=\color{black},
    numberstyle=\tiny\color{gray},
    basicstyle=\ttfamily\footnotesize,
    breaklines=true,
    frame=single,
    rulecolor=\color{gray!50},
    tabsize=4,
    showstringspaces=false,
    captionpos=b,
    numbers=left,
    numbersep=5pt
}
\lstdefinelanguage{diff}{
  basicstyle=\ttfamily\footnotesize,
  morecomment=[f][\color{gray}]{@@},
  morecomment=[f][\color{red}]{-},
  morecomment=[f][\color{green!60!black}]{+},
}

\lstdefinestyle{diffstyle}{
  language=diff,
  showstringspaces=false,
  numbers=none,
  basicstyle=\ttfamily\footnotesize,
  breaklines=true,
  frame=none,
}
\newtcolorbox{mybox}{%
  colback=white,
  colframe=black,
  width=\linewidth,
  arc=0mm,
  boxrule=0.1mm,
  fontupper=\footnotesize,
  before upper={\raggedright},
}

\begin{document}

\title{BitsAI-Fix: LLM-Driven Approach for Automated Lint Error Resolution in Practice}

\author{
    \IEEEauthorblockN{Yuanpeng Li\textsuperscript{*}}
    \IEEEauthorblockA{\textit{ByteDance}\\Hangzhou, China\\liyuanpeng.meta@bytedance.com}
    \\
    \IEEEauthorblockN{Lintao Xie}
    \IEEEauthorblockA{\textit{ByteDance}\\Hangzhou, China\\xielintao@bytedance.com}
    \\
    \IEEEauthorblockN{Yueyan Chen}
    \IEEEauthorblockA{\textit{ByteDance}\\Beijing, China\\chenyueyan@bytedance.com}
    \and
    \IEEEauthorblockN{Qi Long\textsuperscript{*}\textsuperscript{$\ddagger$}}
    \IEEEauthorblockA{\textit{Carnegie Mellon University}\\Pittsburgh, United States\\qilong@andrew.cmu.edu}
    \\
    \IEEEauthorblockN{Xu He}
    \IEEEauthorblockA{\textit{ByteDance}\\Hangzhou, China\\hexu.324@bytedance.com}
    \\
    \IEEEauthorblockN{Wenbo Duan}
    \IEEEauthorblockA{\textit{ByteDance}\\Beijing, China\\duanwenbo@bytedance.com}
    \and
    \IEEEauthorblockN{Zhiyuan Yao\textsuperscript{$\ddagger$}}
    \IEEEauthorblockA{\textit{Zhejiang University}\\Hangzhou, China\\yaozhiyuan@zju.edu.cn}
    \\
    \IEEEauthorblockN{Lu Geng}
    \IEEEauthorblockA{\textit{ByteDance}\\Hangzhou, China\\genglu.32@bytedance.com}
    \and
    \IEEEauthorblockN{Jian Xu\textsuperscript{$^\dagger$}}
    \IEEEauthorblockA{\textit{ByteDance}\\Hangzhou, China\\xujian.1502@bytedance.com}
    \\
    \IEEEauthorblockN{Xin Han}
    \IEEEauthorblockA{\textit{ByteDance}\\Hangzhou, China\\hanxin.hx@bytedance.com}
    
    \thanks{\textsuperscript{*}Equal Contribution}
    \thanks{\textsuperscript{$^\ddagger$}Work done during internship at ByteDance}
    \thanks{\textsuperscript{$^\dagger$}Corresponding Author.}

}
\maketitle

\begin{abstract}
As enterprise codebases continue to grow in scale and complexity, the volume of lint errors far exceeds engineers\verb+'+ manual remediation capacity, leading to continuous accumulation of technical debt and hindered development efficiency. This paper presents BitsAI-Fix, an automated lint error remediation workflow based on Large Language Models (LLMs), designed to address this critical challenge in industrial-scale environments. BitsAI-Fix employs tree-sitter for context expansion and generates search-and-replace format patches through specially trained LLMs, followed by lint scan re-verification to output final remediation results. Additionally, our approach introduces an innovative progressive reinforcement learning (RL) training strategy that can automatically acquire verifiable training data during the project cold-start phase and continuously iterate the model by collecting online samples through feedback after system deployment. Furthermore, we designed a targeted rule-based reward mechanism that combines format rewards and correctness rewards while penalizing redundant modifications. We also propose a “code diff matching" methodology to continuously track online effectiveness. In production deployment at ByteDance, our solution has supported over 5,000 engineers, resolved more than 12,000 static analysis issues, achieved approximately 85\% remediation accuracy, with around 1,000 weekly active adopters. This work demonstrates the practical feasibility of LLM-based code remediation solutions in enterprise environments and serves as a reference for automated code fix in large-scale industrial scenarios.



\end{abstract}

\section{Introduction}

Code fixing is an indispensable yet labor‐intensive task in modern software development.  As enterprise codebases at organizations such as ByteDance expand in scale and complexity, the volume of issues identified by automated Lint scans significantly surpasses engineering capacity for manual remediation.  The sheer volume of warnings and errors discourages timely remediation, consequently leading to accumulated technical debt that degrades code health and impedes feature iteration. To address this scalability challenge, automated code repair has emerged as a promising solution to reduce manual effort and improve code maintenance efficiency. Automated code repair techniques primarily rely on a \verb+"+generate-and-validate\verb+"+ paradigm. These approaches include search-based, template-based and constraint-based methods that generate candidate patches and validate them using test cases~\cite{10.1145/3631974,10.1145/3213846.3213871,10.1145/3131704.3131720}. However, these traditional methods are commonly plagued by challenges such as low repair success rates~\cite{10172803}.

Recent advances in LLMs have demonstrated remarkable capabilities in understanding, generating, and refactoring source code ~\cite{10549472,rozière2024codellamaopenfoundation,nijkamp2023codegenopenlargelanguage}, as well as other code-related tasks such as code review ~\cite{sun2025bitsaicrautomatedcodereview}. Leveraging these capabilities, researchers have begun exploring LLM-based solutions for automated code fixing, with two primary research paradigms emerging. One line of research focuses on enhancing LLMs' intrinsic code repair capabilities through post-training optimization. Supervised fine-tuning (SFT) approaches have demonstrated effectiveness in domain-specific scenarios, with existing literature~\cite{duan2024pdcdmsftroad} showing improved SQL code error correction through targeted model adaptation. RL methods have gained particular traction, where carefully designed reward mechanisms encourage LLMs to explore the generation of correct code. Notably, GRPO~\cite{shao2024deepseekmathpushinglimitsmathematical} has proven that RL with rule-based rewards can yield substantial improvements in code generation quality. In parallel, agent-based methodologies have emerged as a complementary paradigm, particularly within industrial applications~\cite{lee2024unifieddebuggingapproachllmbased}\cite{wang2024executablecodeactionselicit}. These approaches leverage the autonomous planning capabilities of LLMs to address code error resolution through iterative, multi-step processes. Rather than relying solely on single-shot generation, agent-based systems orchestrate multiple LLM interactions to systematically analyze, plan, and execute complex code fix. 

Nevertheless, existing methods encounter several challenges when addressing lint error fixing tasks. 1) Lint error repair demands high accuracy standards, as low fix rates can cause significant user disruption given the massive volume of errors encountered in practice. Current direct model approaches fail to achieve satisfactory accuracy levels~\cite{wen2025fixingfunctionlevelcodegeneration}. While post-training methods such as SFT or RL could potentially improve accuracy, there is limited practical guidance on approach selection, training data acquisition or reward strategy design for RL-based solutions. 2) Agent-based approaches demonstrate substantially improved accuracy but suffer from prohibitive latency and cost overhead when applied to large-scale lint error scenarios, making them overly complex for this specific use case. 3) Lint error detection typically occurs during the merge request (MR) phase, where feedback acquisition remains insufficient. Users commonly resolve issues within their local IDE environments rather than through the repository interface, hindering effective feedback collection and impeding iterative system improvement and data accumulation efforts. To address these limitations, we propose an industrial-grade automated lint error fixing system with an effective model training framework. Specifically, our work makes the following three major contributions:

\begin{itemize}
\item \textbf{Lightweight Lint Error Fix Workflow} 
We propose an streamlined LLM-based workflow rather than an agent-based approach to effectively handle massive volumes of lint errors. Our solution incorporates sophisticated mechanisms including context extraction, result verification, and failure retry strategies to enhance overall repair accuracy. This approach effectively offloads repetitive repair tasks from engineers, accelerates issue resolution, and significantly improves code quality in industrial settings.

   \item  \textbf{Progressive RL Training Strategy} We introduce a novel RL training strategy comprising progressive verifiable data collection and adaptive reward rule design. Our data collection framework operates in two progressive phases: initially cold-start data acquisition through dependency mocking and compilation minimization, followed by continuous feedback collection post-deployment for iterative enhancement. We implement evolving rule-based reward strategies specifically designed for different data acquisition methods at each phase. This progressively refined RL training approach ultimately delivers high-precision minimal patches with non-intrusive repair characteristics through iterative improvement.
    \item \textbf{Effective Assessment\&Iteration Mechanism} 
    We introduce a production-environment assessment mechanism that measures the true acceptance of AI-generated fixes by computing textual and semantic similarity between developer-committed patches and system-suggested repairs. This approach enables the construction of ground-truth labeled datasets from actual development workflows, facilitating continuous self-improvement of the LLM-RL integrated system through real user feedback.


\end{itemize}

 \begin{figure*}[htbp]
  \centering
  \includegraphics[width=0.95\linewidth]{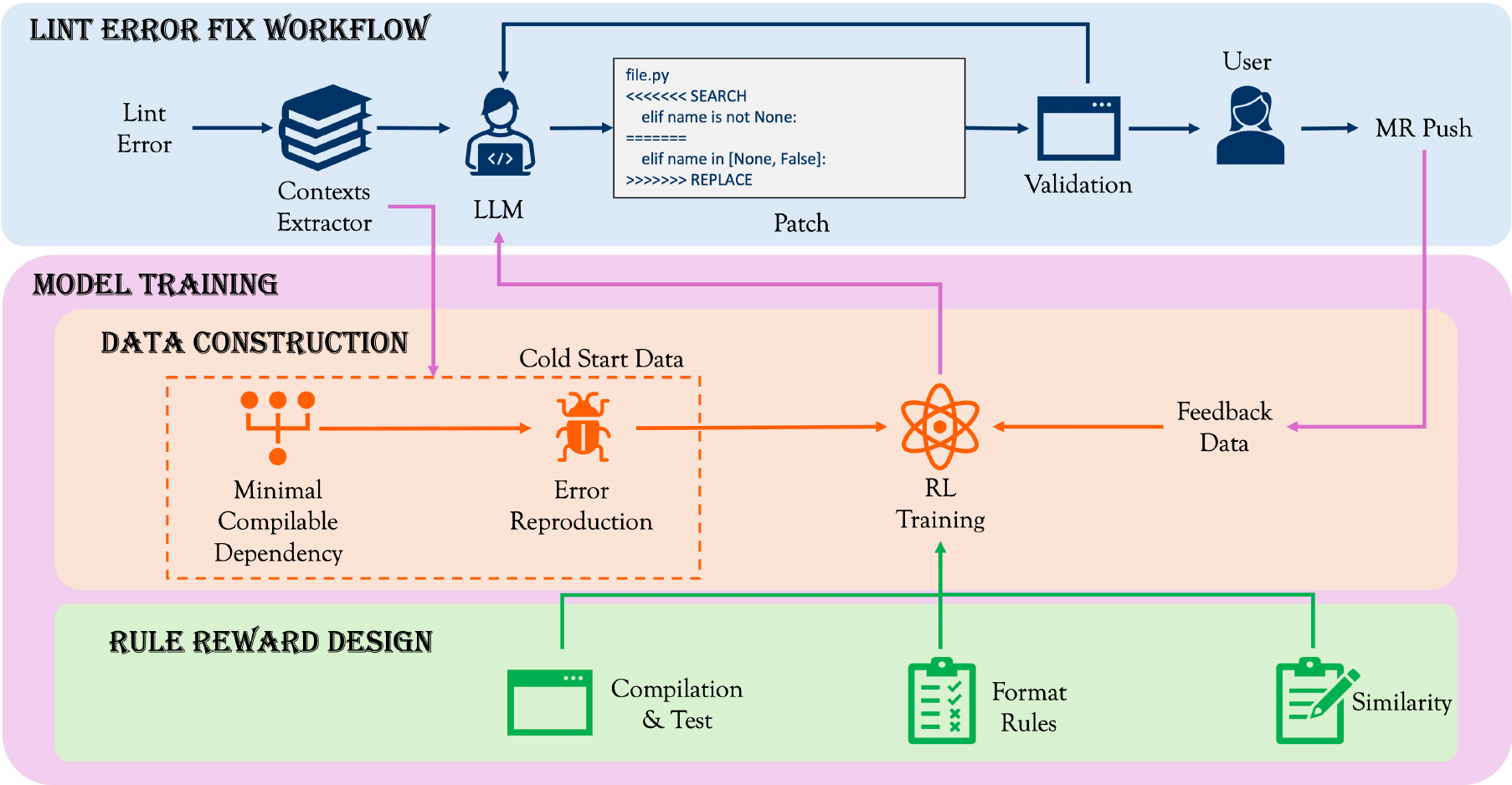}
  \caption{Framework of BitAI-Fix, including Lint Error Fix Workflow and Model Training.}
  \label{fig:framework}
\end{figure*}

To validate these contributions, we conducted a comprehensive large-scale deployment at ByteDance, our system achieves about 85\% repair accuracy and has automatically resolved more than 12,000 issues to date.  This significantly alleviates engineer workload in lint error handling, helps reduce technical debt related to code quality issues, and contributes to improved development efficiency and code maintainability.

\section{Related Work}

\subsection{Traditional Automated Code Fix}
To reduce the large amount of time software engineers spent on fixing code, researchers in the field have dedicated significant efforts to realize automated program repair \cite{10.1145/3631974}. For non-learning based methods, there are several distinct approaches. Firstly, the search-based methods \cite{10.1145/3213846.3213871}\cite{10.1145/3131704.3131720} abstract the repair process into a search problem in a space including all possible patches. Secondly, the constraint-based approaches transform the repair process into a constraint solving problem \cite{10.1145/3106237.3106309}\cite{7816488}. Thirdly, the template-based method utilizes a predefined program fix template to generate patches \cite{10.5555/2486788.2486893, 7476644, 10.1007/s10664-019-09780-z}. Learning-based methods subsequently transformed code fix into a code generation task in text format given buggy codes. This paradigm shift began with the application of Neural Machine Translation (NMT) techniques \cite{10.1145/3340544}, which inspired increasingly more methods to build on training deep learning models \cite{8668043,9284100,10.1145/3395363.3397369}. Recently, The field has experienced a revolutionary advancement with the surge of Large Language Model (LLM) technology, which has pushed the boundaries of Automated Program Repair (APR) significantly \cite{Xia_2023}, demonstrating clear superiority over non-learning based methods.

\subsection{LLM-based Code Fix}
LLMs have brought revolutionary advances to code repair techniques. The LLM-based code fix process typically involves pretraining, SFT and RL. Evaluation work \cite{10232867} shows consistent improvement through pretraining for vulnerability repair tasks, while supervised fine-tuning on APR data effectively enhances code fix capabilities \cite{10232867}\cite{10.1145/3597926.3598135}. For RL, rewards can be derived from similarity comparison with ground truth \cite{wei2025swerladvancingllmreasoning}\cite{islam2024codesecurityvulnerabilityrepair}, compilation results \cite{wang-etal-2022-compilable}, and unit test outcomes \cite{liu2023rltfreinforcementlearningunit}. SWE-Bench\cite{jimenez2024swebench}stands as a challenging benchmark with 2294 tasks from real-world software projects, requiring models to understand complex codebases and generate appropriate fixes across multiple files. To address these challenges, agent-based approaches like SWE-agent \cite{yang2024swe} employ ReAct-style loops to interact with shell commands and specialized tools, while pipeline-based methods such as Agentless \cite{xia2024agentless} decompose the fix workflow into a predetermined sequence of stages—bug reproduction, file localization, code editing, etc.—and solve each step in turn. Evidence suggests LLM-first approaches that allow models to reason and manipulate environments may outperform methods that insert LLMs at specific points in engineered workflows. As the field evolves, hybrid approaches combining agent-based flexibility with pipeline-based structure are emerging to tackle increasingly complex software engineering tasks.

\subsection{Industrial Applications of Code Fix}
In recent years, there has been growing interest in applying LLM for code fix in industrial settings. These applications typically focus on specific vertical scenarios to address real-world software engineering challenges. For instance, CrashFixer \cite{mathai2025crashfixercrashresolutionagent} refined the pipeline of hypothesis, patch generation and patch validation to specialize in addressing kernel crash issues and has achieved notable results in production environments. Similarly, DR.FIX \cite{behrang2025drfixautomaticallyfixingdata} deployed a retrieval system on external race example database for knowledge injection, concentrating on resolving data race problems with impressive efficiency.Furthermore, the Google Team \cite{unknown} has made significant strides by improving the reference retrieval system with LLM assistance in changing and validation processes, demonstrating great success in code migration at Google. These industrial applications highlight the practical value of automated code repair technologies beyond academic research, showing how theoretical advances can be translated into real-world solutions for software development teams.

\section{Methodology}
The overall framework of BitsAI-Fix is illustrated in Figure \ref{fig:framework}, which comprises two primary components: the Lint Error Fixing Workflow, encompassing context extraction, patch generation, and validity verification; and the RL model training module, incorporating verifiable dataset construction and targeted rule-based reward design. Each component is detailed in the following sections.

\subsection{The Lint Error
Fix Workflow}
The online workflow consists of three fundamental modules: Context Extraction, Patch Generation, and Validity Validation. This architecture leverages direct model invocation to produce repair patches, circumventing the computationally intensive multi-round interactions typical of agent-based methodologies, thus rendering it suitable for large-scale lint error repair. Additionally, user-accepted fixes are incorporated as training exemplars for continuous model improvement. Each module is described below:

\textbf{Context Extraction.} For each lint warning (unused imports, missing error checks, style violations, etc.), we employ a two-step context extraction process. Since the location of the issue is already known in lint error scenarios, we can purposefully extract the relevant context. First, we use \texttt{tree\_sitter} to perform function-level expansion, identifying the minimal syntactic unit containing the issue. Second, we extract one layer of dependencies by gathering all directly referenced symbols and their definitions. Our experiments demonstrate that this approach optimally balances repair quality with context efficiency, providing sufficient information for accurate LLM-based fixes while avoiding context overflow.

\textbf{Patch Generation.} We feed the extracted context snippet and original lint message into a specially trained LLM designed for code repair tasks. The model is trained to output search-and-replace–style unified diff format (see Figure ~\ref{fig:search_replace_patch} for details), leveraging domain-specific training to better recognize lint error patterns and produce contextually appropriate solutions. The unified diff format provides three critical benefits: (1) robustness against line number changes during code evolution, (2) minimal token overhead by targeting only modified lines, and (3) standard formatting that enables seamless automatic application upon developer approval.



\begin{figure}[htbp]
\centering
\begin{mybox}
\begin{verbatim}
### project/project_workflow_main_loop.go
<<<<<<< SEARCH
	errg.Go(func() error {
		if err = task.InitIDL(ctx); err != nil {
		      return err
		}
		return nil
	})
=======
	errg.Go(func() error {
		defer func() {
			if r := recover(); r != nil {
					logs.CtxInfo(ctx, 
                  "Recovered from panic: %v", r)
			}
		}()
		return task.InitIDL(ctx)
	})
>>>>>>> REPLACE
\end{verbatim}
\end{mybox}
\caption{Search Replace format for patch generation}
\label{fig:search_replace_patch}
\end{figure}

\textbf{Validity Verification.} Each generated diff undergoes automatic validation before being presented to developers. The system applies the diff to the codebase and performs a lint scan to confirm the original issue is resolved. Failed validations trigger automatic regeneration, with up to 3 retry attempts by default. Successfully validated diffs are then presented in the pull-request interface, where developers can \emph{accept}, \emph{reject}, or \emph{modify} each suggestion. Only explicitly accepted diffs are converted into actual commits.



\subsection{The Strategy of RL Training}
In this section, we present our model training methodology, providing comprehensive descriptions of our data construction pipeline and reward design mechanism. As discussed above, models trained solely through SFT demonstrate insufficient precision in code fix scenarios and fail to achieve minimal, non-intrusive modifications. Therefore, we adopt RL approaches to address these limitations.

The establishment of robust verification frameworks is essential to the efficacy of RL paradigms. In code generation and repair tasks, there has two predominant verification strategies: \textit{(1) Text-matching verification:} this approach assumes the existence of a \verb+"+golden patch\verb+"+ for each buggy sample. The generated patch is compared against the ground-truth label (e.g., via string similarity or edit distance) to compute a reward. While it supports fine-grained control over specific repair patterns, obtaining golden patches is labor-intensive. Notably, such reference data is inherently absent in the initial cold-start scenario. \textit{(2) Execution-based verification:} for execution-based verification, we require that each generated patch compiles successfully and turns a previously failing test case into a passing one—a fail-to-pass validation exactly as in SWE-Bench\cite{jimenez2024swebench}. As evidenced here, this approach obviates the need for ground-truth solutions, enabling reward computation through direct validation of the generated solution's correctness. This paradigm provides a viable pathway for reward signal acquisition in scenarios where reference answers are unavailable.


Given these considerations, we propose a progressive training paradigm: initially leveraging execution-based verification during cold-start when reference data is unavailable, then transitioning to text-matching verification as we accumulate high-quality patches from system deployment. The methodology proceeds as follows:

\subsubsection{Cold-start data construction} In this phase, we primarily focus on constructing executable samples. However, rather than pursuing executable samples with test cases, we creatively approximate executability through compilability and lint tool validation, which has demonstrated remarkably effective results in practice. To ensure sample executability, one feasible approach is to directly execute the entire project. However, industrial-scale projects are typically quite large, requiring substantial time and resources for building, compilation, and execution. This leads to low efficiency in both data acquisition and training. Therefore, we designed a \verb+"+semi-synthetic\verb+"+ data generation process that, through the following three steps, retains only the minimal executable dependencies relevant to the current problem. 
\begin{itemize}
\item \textit{Minimal Dependencies Construction}. 
For entity dependency information within the project, we leverage tree\_sitter for AST parsing whenever possible to obtain \verb+"+actual dependencies\verb+"+. For dependencies on imported third-party packages, given the potential multi-level nesting and complexity of acquiring these packages, we employ a LLM-based simulation approach to construct \verb+"+virtual dependencies\verb+"+.

\item \textit{Issues Reproduction Test.}  
We focus on reproducing the original issues to establish a reliable baseline for validation. This step is primarily aimed at making the sample able to be accurately determined whether its issue has been resolved using a Fail-To-Pass test approach. 
Our primary goal in this step is to establish a reliable baseline for validation by accurately reproducing the original issues. We directly rerun the lint tool on the code with minimal dependencies constructed in the previous step. If the same issue is detected, the reproduction is considered successful. Otherwise, it is deemed a failure. 

 \item \textit{Effective Sample Selection.} 
We ensure the final samples meet three key criteria: accurate verification capability, reasonable difficulty distribution, and balanced knowledge point coverage. Our selection process involves: (1) retaining only samples with successfully constructed minimal dependencies that compile and pass reproduction tests; (2) classifying difficulty using DeepSeek-R1~\cite{deepseekr1} by performing eight repair attempts per sample, where difficulty is determined by the number of successful repairs (samples with all eight successful attempts are excluded as too simple); and (3) selecting up to 30 samples per lint error category while maintaining balanced difficulty distribution across categories.
\end{itemize}

\subsubsection{User Feedback Data Collection} Methods based on execution-based verification can achieve promising results in cold-start reinforcement learning training. However, in our approach, instead of employing test case execution, we utilize compilation and lint tool scanning as an approximation of execution-based verification. This methodology cannot truly guarantee the absolute correctness of the generated code fixes, which to some extent constrains the performance ceiling of RL training: the model tends to generate modifications that pass compilation and lint tool scanning rather than truly correct fixes. Therefore, we still need to obtain samples with \verb+"+golden patches\verb+"+. Evidently, samples from online user feedback satisfy this requirement. In practice, following the small-scale pilot deployment of BitsAI-fix, we began collecting user feedback data and incorporating it into the RL training sample set. Over time, we continuously perform iterative refinement through this process.

\subsubsection{Rule Reward Design}
We follow the GRPO algorithm
for RL and design rule-based rewards conditioned on error type. The GRPO algorithm has its advantage in improving the LLM's robustness when dealing with different groups of preference data. The GRPO objective is formulated as follows \ref{eq:grpo}

\begin{equation}\label{eq:grpo}
\begin{split}
\mathcal{J}_{\text{GRPO}}(\theta) = \; & \mathbb{E}_{\mathbf{q} \sim \mathcal{G}, \, \{o_i\}_{i=1}^G \sim \pi_{\theta_{\text{old}}}(O|\mathbf{q})} \Biggl[ \\
&  \frac{1}{G} \sum_{i=1}^{G} \frac{1}{|o_i|} \sum_{t=1}^{|o_i|}
    \begin{aligned}[t] 
    \min\Biggl(& r_t(\theta)\hat{A}_{i,t}, \\ 
    & \operatorname{clip}(r_t(\theta), 1-\epsilon, 1+\epsilon)\hat{A}_{i,t}\Biggr) 
    \end{aligned}
 \\
& - \beta \, \text{KL}[\pi_{\theta} \| \pi_{\text{ref}}] \Biggr].
\end{split}
\end{equation}
where $\pi_\theta$ denotes the LLM as the current policy model, $\pi_{\theta_{\text{old}}}$ is the previous policy model, and $\mathcal{G}$ includes the aforementioned three data groups. The importance sampling ratio $r_t(\theta)$ is defined as
\begin{equation}
r_t(\theta) = \frac{\pi_\theta(o_{i,t})}{\pi_{\theta_{\text{old}}}(o_{i,t})}
\end{equation}
The reward is designed to be within 0 and 1 point, which is rated based on two main criteria, formatting and correctness. For formatting, the reward follows \ref{eq:format_reward} to deduct points for each mismatch in a penalty manner, where $C$ is the generated code patches, $i$ refers to $i$ patches not in search\&replace format and $j$ refers to $j$ patches not applied. Since we expect one patch generation per error, the patches more than one are regarded as redundant, which is also penalized.
\begin{equation}\label{eq:format_reward}
    r_{f}(C) =
    \begin{cases}
    0.0, & \text{if } C \text{ has correct format} \\
    -0.1\cdot i -0.1\cdot j, & \text{if } C \text{ not S\&R or not applied} \\
    -0.1\cdot k, & \text{if } C \text{ has redundant k patches}
    \end{cases}
\end{equation}
For correctness, the reward follows \ref{eq:compile_reward} and \ref{eq:correctness_reward}, where $C$ is the generated code patches. A compilable precheck $r_{p}(C)$ is conducted before correctness check. If the case is not compilable, the correctness reward is zero directly. Otherwise, for cold-start data, since they do not have ground truth labeling but go through Minimal Dependencies Construction and Issues Reproduction, they are executable and thereby rewarded via fail-to-pass. For feedback data, since they have ground truth labeling, a similarity match between generated code patch and ground truth is conducted according to $F_{\beta}$ \cite{ma2025sorftissueresolvingsubtaskoriented}.
\begin{equation}\label{eq:compile_reward}
    r_{p}(C) =
    \begin{cases}
    0.0, & \text{if } C \text{ is not compilable} \\
    +0.3, & \text{if } C \text{ is compilable}
    \end{cases}
\end{equation}
\begin{equation}\label{eq:correctness_reward}
    r_{c}(C) =
    \begin{cases}
    +0.7, & \text{if } C \text{ is cold-start data and fail-to-pass} \\
    +0.7\cdot F_{\beta}, & \text{if } C \text{ is feedback data}
    \end{cases}
\end{equation}
Finally, the total reward is calculated as \ref{eq:reward_sum}.
\begin{equation}\label{eq:reward_sum}
    r(C) = r_{f}(C) + r_{p}(C) + \pi[r_{p}(C)>0]\cdot r_{c}(C)
\end{equation}

\subsection{The Mechanism of Assessment and Iteration}

Since our approach targets code fixes within the MR scenario, the fix results are presented directly on the MR page. From a product design perspective, we initially evaluated user acceptance of our fix suggestions through click-through tracking on the \verb+"+Adopt\verb+"+ button. However, this approach revealed a significant behavioral discrepancy: users predominantly prefer to modify code within their IDE and submit changes directly, rather than interacting with the MR interface. Consequently, this method yielded extremely limited feedback, rendering it ineffective for accurately measuring the true performance of BitsAI-Fix and insufficient to support our subsequent continuous optimization iterations.

To obtain more accurate user feedback data, we implemented a \verb+"+code diff matching\verb+"+ methodology for evaluation. Specifically, we employ the following approach: we establish the LLM-generated diff patch as the baseline reference, then extract the actual diff after users merge or update their merge requests, and subsequently perform matching between these two artifacts. When the LLM-generated diff patch is completely encompassed within the user's actual submitted diff, we classify the fix as adopted. Conversely, any deviation or partial overlap is categorized as non-adoption.

Through training our reinforcement learning loop based on this feedback mechanism over a two-month iteration period, we generated patches that more closely aligned with developers' authentic coding and remediation practices. Empirically, this evaluation strategy demonstrates superior persuasiveness and reliability compared to conventional copy-or-click proxies, thereby accelerating our iterative model enhancement process.

\section{Experiments}
To comprehensively evaluate the effectiveness of our method, we conducted detailed offline experiments and assessments using the Go programming language as a case study, which serves as the primary programming language at ByteDance.

\subsection{Dataset Construction} To date, we have constructed a training dataset of 20,000 samples, comprising approximately 16,000 cold start samples and 4,000 user feedback samples. These samples are evenly distributed across all 198 lint error categories, ensuring comprehensive enhancement of the model's ability to handle various lint error fix challenges. Additionally, we built a test dataset containing 2,271 samples that similarly covers all 198 lint error categories, with each category containing between 1 and 15 samples.

\subsection{Evaluation Criteria}
To validate the accuracy of repair results and ensure minimal code modifications, we adopt the following two evaluation criteria:
\begin{itemize}
    \item \textbf{Fix Accuracy.} 
    Defined as the proportion of model-generated patches that successfully compile and pass the original lint error scanning when applied to the problematic code. This metric serves as a key indicator for assessing repair effectiveness. The metric is formally expressed as:
\begin{equation}
\label{eq:fix_accuracy}
\text{Fix Accuracy} = \frac{\sum_{i=1}^{N} \mathbb{I}(S_i)}{N}
\end{equation}
where $N$ is the total number of code samples, and $\mathbb{I}(S_i)$ equals 1 if the patch for sample $i$ is successful, and 0 otherwise.
    \item \textbf{Fix Redundancy.}  Defined as the conciseness of the patches generated by the model. For a given problematic code, if the number of search-and-replace blocks generated by the model exceeds the number of lint errors detected, the repair is considered redundant. We measure the overall redundancy rate as the proportion of samples that result in redundant repairs, with a lower rate indicating better performance.This can be mathematically formulated as:
\begin{equation}
\label{eq:fix_redundancy}
\text{Fix Redundancy} = \frac{\sum_{i=1}^{N} \mathbb{I}(P_i > E_i)}{N}
\end{equation}
where $P_i$ is the number of patches generated for sample $i$, and $E_i$ is the number of actual lint errors. The indicator function $\mathbb{I}(P_i > E_i)$ equals 1 if the condition $P_i > E_i$ is true, and 0 otherwise.
\end{itemize}

\subsection{Experiment Setup}

Considering effectiveness, efficiency, and internal compliance requirements, we selected Qwen2.5-Coder-32B~\cite{hui2024qwen25codertechnicalreport} as the base model for fine-tuning.
We fine-tuned it using GRPO-based RL with the verl~\cite{sheng2024hybridflow} framework. The model was trained with a learning rate of $1 \times 10^{-6}$, a global batch size of 256, and a PPO mini-batch size of 128. For KL divergence control, we set the KL loss coefficient to 0.001 using a low-variance formulation. Input sequences were limited to 8,192 tokens with outputs capped at 4,096 tokens, and a rollout of 8 was employed. All training was conducted on Ascend-910B2 NPUs across 8 nodes with 16 NPUs per node.

\begin{figure*}[htbp]
  \centering
  \includegraphics[width=0.95\linewidth]{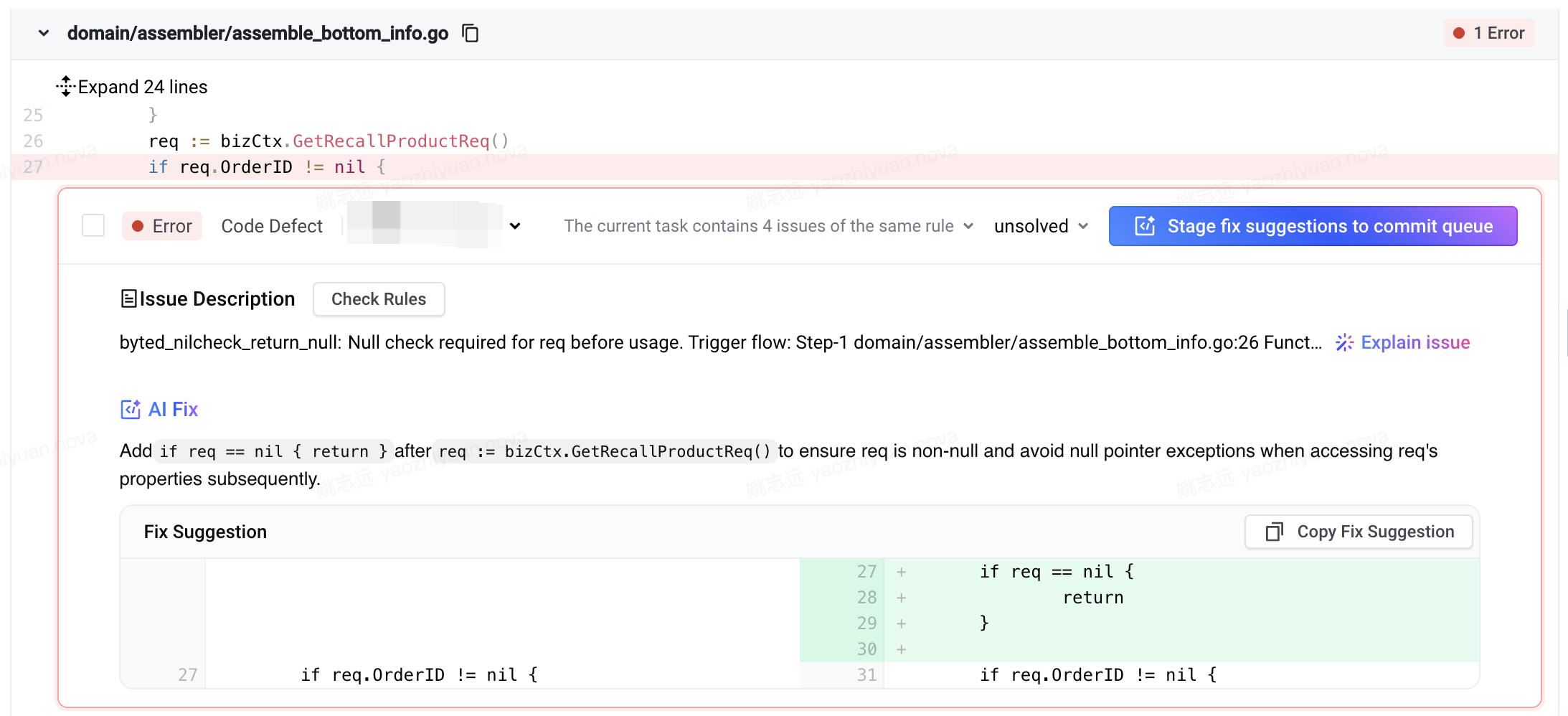}
  \caption{Main interface of the automated code-fix tool embedded in the MR review page.}
  \label{fig:product_page}
\end{figure*}
\subsection{Experiment Results}
We conducted a systematic evaluation comparing the overall accuracy performance of models across distinct training stages, including mainstream base model, models enhanced via SFT, and models further refined through RL optimization. 
\begin{table}[h]
    \centering
    \caption{Performance Comparison of Base, SFT, and RL Models}
    \label{table:experiment_results}
    \begin{tabular}{
    >{\centering\arraybackslash}p{2.5cm}
    >{\centering\arraybackslash}p{1.5cm}
    >{\centering\arraybackslash}p{1.5cm}
    >{\centering\arraybackslash}p{1cm}
}
        \specialrule{1.5pt}{0pt}{0pt}
         & \textbf{Accuracy} & \textbf{Redundancy} & \textbf{Category} \\
        \hline
        \textbf{DeepSeek-R1-0528} &  73.23\%& 12.15\% & Base \\
        \textbf{Doubao1.6-Thinking} &69.70\%  &20.83\%  & Base \\
        \textbf{DeepSeek-V3} &66.18\%  & 17.57\% & Base \\
        \textbf{Qwen2.5-Coder-32B} &53.76\%  &10.00\%  & Base \\
        \textbf{Qwen3-32B} &51.30\%  & 11.71\% & Base \\
        \hline
        \textbf{Qwen2.5-Coder-32B-SFT} &65.48\%  &9.25\%  & SFT \\
        \hline
        \textbf{Qwen2.5-Coder-32B-RL} &84.68\%  &1.72\%  & RL \\
        \specialrule{1.5pt}{0pt}{0pt}
    \end{tabular}
\end{table}
The experimental results presented in Table \ref{table:experiment_results} support the following conclusions: 

\textbf{RL Training Achieves Superior Accuracy.}  Post-training significantly improves base model performance on fix tasks, with both SFT and RL fine-tuning demonstrating substantial gains over the untuned Qwen2.5-Coder-32B model (53.76\%). However, their practical effectiveness differs markedly. While Qwen2.5-Coder-32B-SFT increases accuracy to 65.48\%, this improvement remains insufficient for real-world deployment requirements. In contrast, Qwen2.5-Coder-32B-RL achieves 84.68\% accuracy through GRPO training by optimizing direct objectives and effectively leveraging reward signals to master complex code repair strategies. Remarkably, our 32B RL-trained model outperforms significantly larger models including Doubao-1.6-Thinking and DeepSeek-R1-0528 on fix tasks, demonstrating that targeted RL optimization can compensate for scale limitations while delivering substantial cost savings.

\textbf{RL Training Minimizes Redundant Modifications.}  Untrained models are particularly prone to generating redundant modifications, ultimately resulting in intrusive code repairs that compromise code quality. This challenge persists even among sophisticated large-scale models—both Doubao-1.6-Thinking and DeepSeek-R1-0528 exhibit high redundancy rates without targeted training. Our results demonstrate that task-specific training substantially mitigates this issue: while our Qwen2.5-Coder-32B-SFT model achieves relatively low redundancy compared to untrained models, Qwen2.5-Coder-32B-RL reaches an exceptionally low fix redundancy of 1.72\%. This demonstrates that RL, by directly penalizing unnecessary edits, is more effective at avoiding redundant code modifications.
\subsection{The Impact of Reward Methods}

Regarding the reward design, we compared how different reward schemes affect the final outcome. Based on the comparison in Table II, we observe that the choice of reward scheme significantly affects both repair accuracy and redundancy.

\begin{table}[h]
    \centering
    \caption{Ablation study on reward design strategies}
    \label{table:reward_methods_results}
    \begin{tabular}{
    >{\centering\arraybackslash}p{4.5cm}
    >{\centering\arraybackslash}p{1.5cm}
    >{\centering\arraybackslash}p{1.5cm}
}
        \specialrule{1.5pt}{0pt}{0pt}
         & \textbf{Accuracy} & \textbf{Redundancy}\\
        \hline
        \textbf{Binary Pass/Fail Reward} & 82.78\% & 6.21\%\\
        \textbf{Graded Reward} &84.02\%  & 7.62\%\\
        \textbf{Graded Reward with Redundancy Penalty} & 84.68\%  & 1.72\% \\
        \specialrule{1.5pt}{0pt}{0pt}
    \end{tabular}
\end{table}
Starting with a simple \textbf{Binary Pass/Fail Reward} (1 point if the issue is fixed; 0 otherwise), the model achieves 82.78\% accuracy with 6.21\% redundancy, serving as a solid baseline. When we introduce a \textbf{Graded Reward} system that splits into compilation and correctness-verification components (0 points for compilation failure, 0.3 points for successful compilation, 1 point for complete fix), accuracy rises to 84.02\% but redundancy increases to 7.62\%. This represents only a 1.24-point gain in accuracy alongside a 1.41-point rise in redundancy, suggesting that rewarding compilation success alone yields limited behavioral change without explicit penalties.

However, augmenting the \textbf{Graded Reward With Redundancy Penalty} (subtracting 0.1 points for each redundant search-and-replace block) produces markedly different results. This approach further boosts accuracy to 84.68\% while sharply reducing redundancy to 1.72\%. These findings demonstrate that penalizing redundant search-and-replace blocks effectively guides the model toward concise and precise fixes without sacrificing overall performance.

\section{Industrial Deployment and Empirical Evaluation}
This chapter presents a comprehensive evaluation of BitsAI-Fix in real-world production environments, covering the system implementation, large-scale industrial deployment, online product performance analysis, and representative case studies.

\subsection{Implementation of BitsAI-Fix}
\label{sec:full_rollout_metrics}
Developers can easily enable and utilize BitsAI-Fix features as demonstrated in Figure ~\ref{fig:product_page}. The system operates by accepting a lint error as \textbf{input}, enabling users to examine the specific rule violation details through the “View Rules" button located adjacent to the “Issue Description" tab. The system then generates two complementary \textbf{outputs}: fix recommendations that explain the rationale behind suggested changes, and concrete repair code prominently displayed on the main interface with a clear side-by-side comparison between the original and corrected versions.

Upon reviewing the system's fix results, users have two convenient options for proceeding. If they approve of the suggested corrections, they may select the \textbf{“Stage Fix Suggestion to Commit Queue"} button to seamlessly prepare the changes for subsequent version control commits. Alternatively, users can employ the \textbf{“Copy Fix Suggestion"} button to transfer the corrected code to their clipboard, allowing them to manually implement the repair within their local IDE environment. This dual-option approach accommodates different developer workflows and preferences for code integration.

\subsection{Large-Scale Industrial Deployment}
    
The deployment of BitsAI-Fix followed a carefully structured two-phase approach to ensure system reliability and performance optimization.

\textbf{Before May: Small-Scale Pilot} This phase commenced limited canary deployment to conduct initial system validation and workflow verification. It also corresponds to the cold-start phase in our progressive training approach. After training the model with cold-start data, we launched this phase to obtain the first batch of online feedback data for further model performance improvement.

\textbf{After May: Full-Scale Rollout} After achieving predefined performance metrics during the pilot, we expanded deployment across essentially all active repositories. The large-scale rollout enabled us to collect extensive positive and negative samples, further enriching our training dataset. This comprehensive implementation demonstrated exceptional system stability and high fix success rates in production.

 \begin{figure}[ht]
  \centering
  \includegraphics[width=0.95\linewidth]{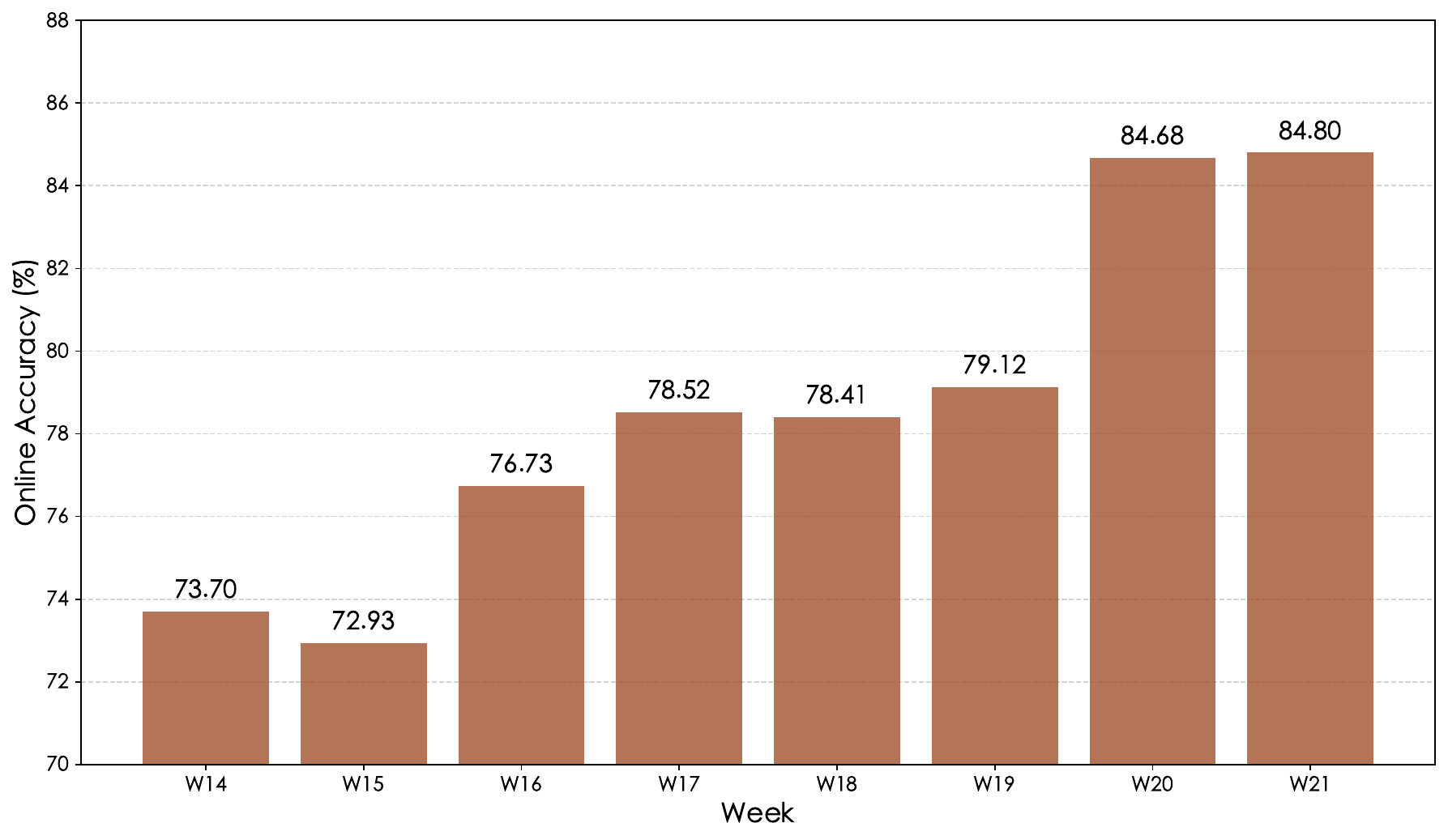}
  \caption{Performance improvement of model optimized using user-feedback dataset across weeks}
  \label{fig:online_acc}
\end{figure}



\subsection{Online Product Performance}
\label{subsec:innovative-evaluation}

\textbf{Online Accuracy Performance} During our rapid iteration period (before May), we implemented a weekly model iteration process leveraging user feedback data. Figure \ref{fig:online_acc} presents senior engineers' manual annotations of real production lint error fixes, demonstrating substantial accuracy improvements from 73.70\% to 84.80\%. Subsequently, we shifted focus to targeted optimizations addressing specific problem cases encountered in production environments. This significant enhancement validates the effectiveness of our approach in identifying recurring error patterns and continuously adapting to evolving developer needs.

\begin{figure}[ht]
  \centering
  \includegraphics[width=0.95\linewidth]{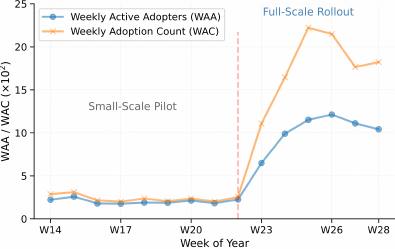}
  \caption{Trend of Weekly Active Adopters and Weekly Adoption Count (Red line indicates Full-Scale Rollout start in May)}
  \label{fig:wau_trend}
\end{figure}

\textbf{User Adoption Metrics}  
Since full-scale implementation, engineering teams have incorporated AI-generated code fixes for more than \textbf{12{,}000} distinct code issues, with participation from over \textbf{5{,}000} unique developers. Recent telemetry indicates sustained adoption rates exceeding \textbf{2{,}000} weekly fix implementations, with approximately \textbf{1{,}000} Weekly Active Adopters (WAA), defined as developers who implement $\ge{1}$ fix per week. Figure~\ref{fig:wau_trend} presents the progressive growth trajectory of Weekly Active Adopters while also tracking the Weekly Adopter Count (WAC), representing the total number of code issues addressed each week.

\subsection{Case Study}
\label{sec:case-study}

In this subsection, we present three representative cases drawn from real‐world issue repositories. Each case illustrates a typical defect pattern, the corresponding automatic fix produced by our system, and the user’s feedback.

\textbf{Case1: Missing recover in Goroutine} In this case, the scheduler started a new goroutine to execute its main loop. However, due to the lack of a recover mechanism within the goroutine, any panic occurring in the main loop would crash the entire scheduler process, affecting system stability. To address this issue, BitsAI-Fix automatically inserted panic capture and logging logic when starting the goroutine, as shown in Figure~\ref{fig:byted_goroutine_recover}. As a result, exceptions in the main loop can be promptly caught and logged without causing the scheduler process to crash, thereby improving system robustness.

\begin{figure*}[htbp]
\centering
\begin{mybox}
\begin{verbatim}
@@ -65,7 +65,14 @@
 // Start begins the scheduler's main loop.
 func (s *SchedulerImpl) Start(ctx context.Context) {
   s.ticker = time.NewTicker(s.interval)
-  go s.run(ctx)
+  go func() {
+    defer func() {
+      if r := recover(); r != nil {
+        logs.CtxError(ctx, "Recovered from panic in SchedulerImpl.run: %v", r)
+      }
+    }()
+    s.run(ctx)
+  }()
   logs.CtxInfo(ctx, "Scheduler started with interval: %v", s.interval)
}
\end{verbatim}
\end{mybox}
\caption{BitsAI-Fix Adding Panic Recovery Mechanism to Goroutine}
\label{fig:byted_goroutine_recover}
\end{figure*}

\textbf{Case2: Unsafe Type Assertion} In this case, the \emph{Execute} method of the finite state machine contained an unchecked type assertion. If the passed \emph{eventCtx} is not actually of type \emph{ShipmentContext}, the program would trigger a panic at runtime, thereby affecting system stability. To resolve this issue, BitsAI-Fix replaced the original forced type assertion with a form that includes type checking, and explicitly returns an error when the assertion fails. This way, even when types don't match, the situation can be handled gracefully through the error handling mechanism without causing exceptional crashes. The fix solution is shown in Figure~\ref{fig:byted_interface_check_golintx}.

\begin{figure*}[htbp]
\centering
\begin{mybox}
\begin{verbatim}
@@ -137,7 +137,10 @@
 type FMReturnShipmentCollectionAllocatedAction struct{}
 
 func (a *FMReturnShipmentCollectionAllocatedAction) Execute(eventCtx EventContext) error {
-    shipmentCtx := eventCtx.(ShipmentContext)
+    shipmentCtx, ok := eventCtx.(ShipmentContext)
+    if !ok {
+        return fmt.Errorf("type assertion failed: eventCtx is not of type ShipmentContext")
+    }
     shipment := shipmentCtx.Shipment
     logger.WithContext(context.Background()).Infof(
         "Executing FMReturnShipmentCollectionAllocatedAction: %v, id: %d",
         shipmentCtx, shipment.ID)
 }
\end{verbatim}
\end{mybox}
\caption{BitsAI-Fix Converting Unsafe Type Assertion to Safe Checked Form}
\label{fig:byted_interface_check_golintx}
\end{figure*}

\textbf{Case3: Over‐eager Integer‐Overflow} In this case, a static analysis warning regarding potential integer overflow from \emph{strconv.Atoi} result conversion to \emph{int16/32} triggered an automatic patch. The system replaced \emph{strconv.Atoi} with \emph{strconv.ParseInt} and cast the result to \emph{int32} to address the overflow concern. However, the fix exhibited over-eagerness by modifying unrelated code regions beyond the scope of the original warning. As shown in Figure~\ref{fig:G109}, the patch introduced additional changes including error message modifications and extra return statements, creating redundancy and confusion in the codebase. The developer deemed these excessive modifications unnecessary and rejected the suggestion, highlighting the importance of maintaining focused, minimal changes that directly address the identified issue without introducing unintended side effects.

\begin{figure*}[htbp]
\centering
\begin{mybox}
\begin{verbatim}
@@ -23,7 +23,9 @@
 	thirdPartySticker *sticker_common.ThirdPartySticker) *sticker_common.StickerItem {
-	normalImageWidth, err := strconv.Atoi(thirdPartySticker.NormalImage.Width)
+	normalImageWidth64, err := strconv.ParseInt(thirdPartySticker.NormalImage.Width, 10, 32)
+	normalImageWidth := int32(normalImageWidth64)
 	if err != nil {
-		logs.CtxError(ctx, "invalid normalImageWidth", err, "Width", thirdPartySticker.NormalImage.Width)
+		logs.CtxError(ctx, "invalid normal image width", err, "Width", thirdPartySticker.NormalImage.Width)
+		return nil
 	}
	normalImageHeight64, err := strconv.ParseInt(thirdPartySticker.NormalImage.Height, 10, 32)
	normalImageHeight := int32(normalImageHeight64)
\end{verbatim}
\end{mybox}
\caption{BitsAI-Fix Over-eager Patch with Excessive Modifications}
\label{fig:G109}
\end{figure*}

Cases1 and Case2 demonstrate that targeted, minimal edits—properly scoped to the relevant code—are highly effective and readily accepted by developers. Case3 reveals the risk of over‐generalizing a fix and unintentionally altering unrelated code. Based on this observation, we refined our patch‐generation strategy to constrain edits strictly to the locations implicated by the reported issue.

\section{Lessons Learned and Practical Insights}
In deploying BitsAI-Fix at ByteDance scale, we distilled three core lessons that can guide other organizations in building reliable, high-impact automated code-fix systems:
 
 
\subsubsection{Lightweight Solutions for Lint Error Fix Scenarios}
While much current research focuses on leveraging large-scale models with complex CodeAgent frameworks for code repair, our work demonstrates that for lint error scenarios—where error locations are well-defined and repair volumes are substantial—smaller LLMs combined with simplified workflows can achieve impressive results while being more practical and cost-effective. This finding offers valuable insights to the community that not all code repair tasks require heavyweight solutions.

\subsubsection{Progressive Training Method for Industrial Code Fix}
We developed a comprehensive RL pipeline covering the complete data lifecycle: from cold-start semi-synthetic dataset construction with executable verification, to post-deployment feedback loops, and carefully designed rule-based rewards. This approach improved our fix rate from 53.76\% to 84.68\% while reducing redundancy from nearly 10\% to 1.72\%. Both our data construction strategy and rule design methodology provide valuable references for industrial applications, demonstrating a scalable framework for deploying code fix systems in production environments. 
\subsubsection{Effective Evaluation Framework for Industrial Deployment}
We developed an effective evaluation methodology for code fix performance that addresses real-world deployment challenges. While automated code fixes on platforms like GitHub are common, users naturally prefer performing fixes directly in their IDEs, making system effectiveness evaluation difficult. Our proposed approach using code change block matching effectively resolves this issue, providing valuable insights for industrial implementations where traditional evaluation metrics fall short due to user behavior patterns.

\section{Conclusion and Future Work}

This paper proposes an industrial-grade workflow for automatically repairing Lint errors based on large language model---BitsAI-Fix, which efficiently addresses the daily Lint scanning issues in enterprise-level codebases and significantly reduces the workload of software engineers. To tackle the limitations of existing pre-trained LLMs in this scenario, we employ reinforcement learning strategies to enhance the model's code repair capabilities. We emphasize the importance of progressive and verifiable data in model training, which includes constructing an executable and verifiable cold-start dataset as well as a continuously evolving, user-annotated feedback dataset. Different rule-based reward mechanisms are designed according to the source of the data. We validate the effectiveness of our approach in real production environments, achieving high fix accuracy, and successfully demonstrate its scalability and practical value through deployment in enterprise-scale software development setting


Although BitsAI-Fix has demonstrated satisfactory accuracy in lint error correction, the system still has certain limitations that warrant further improvement in future research.First, the current system's repair scope is primarily limited to lint errors. To build a more comprehensive code repair solution, we plan to extend the system's capabilities to a broader range of error types, thereby enhancing its applicability in real-world development scenarios.Second, the code context information provided to the model by the existing system is relatively limited, covering only single-layer function nesting structures. This limitation in context scope may adversely affect repair accuracy. To address this, we propose to introduce a dynamic context supplementation mechanism in subsequent work that adaptively adjusts the coverage of context information based on specific repair requirements.Looking ahead, with the continuous development of related technologies and deepening application practices, BitsAI-Fix and its future versions are expected to play an increasingly important role in building high-quality, highly reliable, and highly maintainable software systems, thereby enabling developers to dedicate more effort to creative core development work.




\bibliographystyle{IEEEtran}
\bibliography{references}

\begin{thebibliography}{10}
\providecommand{\url}[1]{#1}
\csname url@samestyle\endcsname
\providecommand{\newblock}{\relax}
\providecommand{\bibinfo}[2]{#2}
\providecommand{\BIBentrySTDinterwordspacing}{\spaceskip=0pt\relax}
\providecommand{\BIBentryALTinterwordstretchfactor}{4}
\providecommand{\BIBentryALTinterwordspacing}{\spaceskip=\fontdimen2\font plus
\BIBentryALTinterwordstretchfactor\fontdimen3\font minus \fontdimen4\font\relax}
\providecommand{\BIBforeignlanguage}[2]{{%
\expandafter\ifx\csname l@#1\endcsname\relax
\typeout{** WARNING: IEEEtran.bst: No hyphenation pattern has been}%
\typeout{** loaded for the language `#1'. Using the pattern for}%
\typeout{** the default language instead.}%
\else
\language=\csname l@#1\endcsname
\fi
#2}}
\providecommand{\BIBdecl}{\relax}
\BIBdecl

\bibitem{10.1145/3631974}
\BIBentryALTinterwordspacing
Q.~Zhang, C.~Fang, Y.~Ma, W.~Sun, and Z.~Chen, ``A survey of learning-based automated program repair,'' \emph{ACM Trans. Softw. Eng. Methodol.}, vol.~33, no.~2, Dec. 2023. [Online]. Available: \url{https://doi.org/10.1145/3631974}
\BIBentrySTDinterwordspacing

\bibitem{10.1145/3213846.3213871}
\BIBentryALTinterwordspacing
J.~Jiang, Y.~Xiong, H.~Zhang, Q.~Gao, and X.~Chen, ``Shaping program repair space with existing patches and similar code,'' in \emph{Proceedings of the 27th ACM SIGSOFT International Symposium on Software Testing and Analysis}, ser. ISSTA 2018.\hskip 1em plus 0.5em minus 0.4em\relax New York, NY, USA: Association for Computing Machinery, 2018, p. 298–309. [Online]. Available: \url{https://doi.org/10.1145/3213846.3213871}
\BIBentrySTDinterwordspacing

\bibitem{10.1145/3131704.3131720}
\BIBentryALTinterwordspacing
Y.~Wang, Y.~Chen, B.~Shen, and H.~Zhong, ``Crsearcher: Searching code database for repairing bugs,'' in \emph{Proceedings of the 9th Asia-Pacific Symposium on Internetware}, ser. Internetware '17.\hskip 1em plus 0.5em minus 0.4em\relax New York, NY, USA: Association for Computing Machinery, 2017. [Online]. Available: \url{https://doi.org/10.1145/3131704.3131720}
\BIBentrySTDinterwordspacing

\bibitem{10172803}
C.~S. Xia, Y.~Wei, and L.~Zhang, ``Automated program repair in the era of large pre-trained language models,'' in \emph{2023 IEEE/ACM 45th International Conference on Software Engineering (ICSE)}, 2023, pp. 1482--1494.

\bibitem{10549472}
X.~Du, M.~Liu, K.~Wang, H.~Wang, J.~Liu, Y.~Chen, J.~Feng, C.~Sha, X.~Peng, and Y.~Lou, ``Evaluating large language models in class-level code generation,'' in \emph{2024 IEEE/ACM 46th International Conference on Software Engineering (ICSE)}, 2024, pp. 982--994.

\bibitem{rozière2024codellamaopenfoundation}
\BIBentryALTinterwordspacing
B.~Rozière, J.~Gehring, F.~Gloeckle, S.~Sootla, I.~Gat, X.~E. Tan, Y.~Adi, J.~Liu, R.~Sauvestre, T.~Remez, J.~Rapin, A.~Kozhevnikov, I.~Evtimov, J.~Bitton, M.~Bhatt, C.~C. Ferrer, A.~Grattafiori, W.~Xiong, A.~Défossez, J.~Copet, F.~Azhar, H.~Touvron, L.~Martin, N.~Usunier, T.~Scialom, and G.~Synnaeve, ``Code llama: Open foundation models for code,'' 2024. [Online]. Available: \url{https://arxiv.org/abs/2308.12950}
\BIBentrySTDinterwordspacing

\bibitem{nijkamp2023codegenopenlargelanguage}
\BIBentryALTinterwordspacing
E.~Nijkamp, B.~Pang, H.~Hayashi, L.~Tu, H.~Wang, Y.~Zhou, S.~Savarese, and C.~Xiong, ``Codegen: An open large language model for code with multi-turn program synthesis,'' 2023. [Online]. Available: \url{https://arxiv.org/abs/2203.13474}
\BIBentrySTDinterwordspacing

\bibitem{sun2025bitsaicrautomatedcodereview}
\BIBentryALTinterwordspacing
T.~Sun, J.~Xu, Y.~Li, Z.~Yan, G.~Zhang, L.~Xie, L.~Geng, Z.~Wang, Y.~Chen, Q.~Lin, W.~Duan, and K.~Sui, ``Bitsai-cr: Automated code review via llm in practice,'' 2025. [Online]. Available: \url{https://arxiv.org/abs/2501.15134}
\BIBentrySTDinterwordspacing

\bibitem{duan2024pdcdmsftroad}
\BIBentryALTinterwordspacing
Y.~Duan, Y.~Yu, X.~Zhao, Y.~Wu, and W.~Liu, ``Pdc \& dm-sft: A road for llm sql bug-fix enhancing,'' 2024. [Online]. Available: \url{https://arxiv.org/abs/2411.06767}
\BIBentrySTDinterwordspacing

\bibitem{shao2024deepseekmathpushinglimitsmathematical}
\BIBentryALTinterwordspacing
Z.~Shao, P.~Wang, Q.~Zhu, R.~Xu, J.~Song, X.~Bi, H.~Zhang, M.~Zhang, Y.~K. Li, Y.~Wu, and D.~Guo, ``Deepseekmath: Pushing the limits of mathematical reasoning in open language models,'' 2024. [Online]. Available: \url{https://arxiv.org/abs/2402.03300}
\BIBentrySTDinterwordspacing

\bibitem{lee2024unifieddebuggingapproachllmbased}
\BIBentryALTinterwordspacing
C.~Lee, C.~S. Xia, L.~Yang, J.~tse Huang, Z.~Zhu, L.~Zhang, and M.~R. Lyu, ``A unified debugging approach via llm-based multi-agent synergy,'' 2024. [Online]. Available: \url{https://arxiv.org/abs/2404.17153}
\BIBentrySTDinterwordspacing

\bibitem{wang2024executablecodeactionselicit}
\BIBentryALTinterwordspacing
X.~Wang, Y.~Chen, L.~Yuan, Y.~Zhang, Y.~Li, H.~Peng, and H.~Ji, ``Executable code actions elicit better llm agents,'' 2024. [Online]. Available: \url{https://arxiv.org/abs/2402.01030}
\BIBentrySTDinterwordspacing

\bibitem{wen2025fixingfunctionlevelcodegeneration}
\BIBentryALTinterwordspacing
H.~Wen, Y.~Zhu, C.~Liu, X.~Ren, W.~Du, and M.~Yan, ``Fixing function-level code generation errors for foundation large language models,'' 2025. [Online]. Available: \url{https://arxiv.org/abs/2409.00676}
\BIBentrySTDinterwordspacing

\bibitem{10.1145/3106237.3106309}
\BIBentryALTinterwordspacing
X.-B.~D. Le, D.-H. Chu, D.~Lo, C.~Le~Goues, and W.~Visser, ``S3: syntax- and semantic-guided repair synthesis via programming by examples,'' in \emph{Proceedings of the 2017 11th Joint Meeting on Foundations of Software Engineering}, ser. ESEC/FSE 2017.\hskip 1em plus 0.5em minus 0.4em\relax New York, NY, USA: Association for Computing Machinery, 2017, p. 593–604. [Online]. Available: \url{https://doi.org/10.1145/3106237.3106309}
\BIBentrySTDinterwordspacing

\bibitem{7816488}
X.-B.~D. Le, Q.~L. Le, D.~Lo, and C.~Le~Goues, ``Enhancing automated program repair with deductive verification,'' in \emph{2016 IEEE International Conference on Software Maintenance and Evolution (ICSME)}, 2016, pp. 428--432.

\bibitem{10.5555/2486788.2486893}
D.~Kim, J.~Nam, J.~Song, and S.~Kim, ``Automatic patch generation learned from human-written patches,'' in \emph{Proceedings of the 2013 International Conference on Software Engineering}, ser. ICSE '13.\hskip 1em plus 0.5em minus 0.4em\relax IEEE Press, 2013, p. 802–811.

\bibitem{7476644}
X.~B.~D. Le, D.~Lo, and C.~Le~Goues, ``History driven program repair,'' in \emph{2016 IEEE 23rd International Conference on Software Analysis, Evolution, and Reengineering (SANER)}, vol.~1, 2016, pp. 213--224.

\bibitem{10.1007/s10664-019-09780-z}
\BIBentryALTinterwordspacing
A.~Koyuncu, K.~Liu, T.~F. Bissyand\'{e}, D.~Kim, J.~Klein, M.~Monperrus, and Y.~Le~Traon, ``Fixminer: Mining relevant fix patterns for automated program repair,'' \emph{Empirical Softw. Engg.}, vol.~25, no.~3, p. 1980–2024, May 2020. [Online]. Available: \url{https://doi.org/10.1007/s10664-019-09780-z}
\BIBentrySTDinterwordspacing

\bibitem{10.1145/3340544}
\BIBentryALTinterwordspacing
M.~Tufano, C.~Watson, G.~Bavota, M.~D. Penta, M.~White, and D.~Poshyvanyk, ``An empirical study on learning bug-fixing patches in the wild via neural machine translation,'' \emph{ACM Trans. Softw. Eng. Methodol.}, vol.~28, no.~4, Sep. 2019. [Online]. Available: \url{https://doi.org/10.1145/3340544}
\BIBentrySTDinterwordspacing

\bibitem{8668043}
M.~White, M.~Tufano, M.~Martínez, M.~Monperrus, and D.~Poshyvanyk, ``Sorting and transforming program repair ingredients via deep learning code similarities,'' in \emph{2019 IEEE 26th International Conference on Software Analysis, Evolution and Reengineering (SANER)}, 2019, pp. 479--490.

\bibitem{9284100}
Y.~Li, S.~Wang, and T.~N. Nguyen, ``Dlfix: Context-based code transformation learning for automated program repair,'' in \emph{2020 IEEE/ACM 42nd International Conference on Software Engineering (ICSE)}, 2020, pp. 602--614.

\bibitem{10.1145/3395363.3397369}
\BIBentryALTinterwordspacing
T.~Lutellier, H.~V. Pham, L.~Pang, Y.~Li, M.~Wei, and L.~Tan, ``Coconut: combining context-aware neural translation models using ensemble for program repair,'' in \emph{Proceedings of the 29th ACM SIGSOFT International Symposium on Software Testing and Analysis}, ser. ISSTA 2020.\hskip 1em plus 0.5em minus 0.4em\relax New York, NY, USA: Association for Computing Machinery, 2020, p. 101–114. [Online]. Available: \url{https://doi.org/10.1145/3395363.3397369}
\BIBentrySTDinterwordspacing

\bibitem{Xia_2023}
\BIBentryALTinterwordspacing
C.~S. Xia, Y.~Wei, and L.~Zhang, ``Automated program repair in the era of large pre-trained language models,'' in \emph{2023 IEEE/ACM 45th International Conference on Software Engineering (ICSE)}.\hskip 1em plus 0.5em minus 0.4em\relax IEEE, May 2023, p. 1482–1494. [Online]. Available: \url{http://dx.doi.org/10.1109/ICSE48619.2023.00129}
\BIBentrySTDinterwordspacing

\bibitem{10232867}
Q.~Zhang, C.~Fang, B.~Yu, W.~Sun, T.~Zhang, and Z.~Chen, ``Pre-trained model-based automated software vulnerability repair: How far are we?'' \emph{IEEE Transactions on Dependable and Secure Computing}, vol.~21, no.~4, pp. 2507--2525, 2024.

\bibitem{10.1145/3597926.3598135}
\BIBentryALTinterwordspacing
Y.~Wu, N.~Jiang, H.~V. Pham, T.~Lutellier, J.~Davis, L.~Tan, P.~Babkin, and S.~Shah, ``How effective are neural networks for fixing security vulnerabilities,'' in \emph{Proceedings of the 32nd ACM SIGSOFT International Symposium on Software Testing and Analysis}, ser. ISSTA 2023.\hskip 1em plus 0.5em minus 0.4em\relax New York, NY, USA: Association for Computing Machinery, 2023, p. 1282–1294. [Online]. Available: \url{https://doi.org/10.1145/3597926.3598135}
\BIBentrySTDinterwordspacing

\bibitem{wei2025swerladvancingllmreasoning}
\BIBentryALTinterwordspacing
Y.~Wei, O.~Duchenne, J.~Copet, Q.~Carbonneaux, L.~Zhang, D.~Fried, G.~Synnaeve, R.~Singh, and S.~I. Wang, ``Swe-rl: Advancing llm reasoning via reinforcement learning on open software evolution,'' 2025. [Online]. Available: \url{https://arxiv.org/abs/2502.18449}
\BIBentrySTDinterwordspacing

\bibitem{islam2024codesecurityvulnerabilityrepair}
\BIBentryALTinterwordspacing
N.~T. Islam, M.~B. Karkevandi, and P.~Najafirad, ``Code security vulnerability repair using reinforcement learning with large language models,'' 2024. [Online]. Available: \url{https://arxiv.org/abs/2401.07031}
\BIBentrySTDinterwordspacing

\bibitem{wang-etal-2022-compilable}
\BIBentryALTinterwordspacing
X.~Wang, Y.~Wang, Y.~Wan, F.~Mi, Y.~Li, P.~Zhou, J.~Liu, H.~Wu, X.~Jiang, and Q.~Liu, ``Compilable neural code generation with compiler feedback,'' in \emph{Findings of the Association for Computational Linguistics: ACL 2022}, S.~Muresan, P.~Nakov, and A.~Villavicencio, Eds.\hskip 1em plus 0.5em minus 0.4em\relax Dublin, Ireland: Association for Computational Linguistics, May 2022, pp. 9--19. [Online]. Available: \url{https://aclanthology.org/2022.findings-acl.2/}
\BIBentrySTDinterwordspacing

\bibitem{liu2023rltfreinforcementlearningunit}
\BIBentryALTinterwordspacing
J.~Liu, Y.~Zhu, K.~Xiao, Q.~Fu, X.~Han, W.~Yang, and D.~Ye, ``Rltf: Reinforcement learning from unit test feedback,'' 2023. [Online]. Available: \url{https://arxiv.org/abs/2307.04349}
\BIBentrySTDinterwordspacing

\bibitem{jimenez2024swebench}
\BIBentryALTinterwordspacing
C.~E. Jimenez, J.~Yang, A.~Wettig, S.~Yao, K.~Pei, O.~Press, and K.~R. Narasimhan, ``{SWE}-bench: Can language models resolve real-world github issues?'' in \emph{The Twelfth International Conference on Learning Representations}, 2024. [Online]. Available: \url{https://openreview.net/forum?id=VTF8yNQM66}
\BIBentrySTDinterwordspacing

\bibitem{yang2024swe}
J.~Yang, C.~E. Jimenez, A.~Wettig, K.~Lieret, S.~Yao, K.~Narasimhan, and O.~Press, ``Swe-agent: Agent-computer interfaces enable automated software engineering,'' \emph{Advances in Neural Information Processing Systems}, vol.~37, pp. 50\,528--50\,652, 2024.

\bibitem{xia2024agentless}
C.~S. Xia, Y.~Deng, S.~Dunn, and L.~Zhang, ``Agentless: Demystifying llm-based software engineering agents,'' \emph{arXiv preprint arXiv:2407.01489}, 2024.

\bibitem{mathai2025crashfixercrashresolutionagent}
\BIBentryALTinterwordspacing
A.~Mathai, C.~Huang, S.~Ma, J.~Kim, H.~Mitchell, A.~Nogikh, P.~Maniatis, F.~Ivančić, J.~Yang, and B.~Ray, ``Crashfixer: A crash resolution agent for the linux kernel,'' 2025. [Online]. Available: \url{https://arxiv.org/abs/2504.20412}
\BIBentrySTDinterwordspacing

\bibitem{behrang2025drfixautomaticallyfixingdata}
\BIBentryALTinterwordspacing
F.~Behrang, Z.~Zhang, G.-V. Saioc, P.~Liu, and M.~Chabbi, ``Dr.fix: Automatically fixing data races at industry scale,'' 2025. [Online]. Available: \url{https://arxiv.org/abs/2504.15637}
\BIBentrySTDinterwordspacing

\bibitem{unknown}
C.~Ziftci, S.~Nikolov, A.~Sjövall, B.~Kim, D.~Codecasa, and M.~Kim, ``Migrating code at scale with llms at google,'' 04 2025.

\bibitem{deepseekr1}
\BIBentryALTinterwordspacing
DeepSeek-AI, ``Deepseek-r1: Incentivizing reasoning capability in llms via reinforcement learning,'' 2025. [Online]. Available: \url{https://arxiv.org/abs/2501.12948}
\BIBentrySTDinterwordspacing

\bibitem{ma2025sorftissueresolvingsubtaskoriented}
\BIBentryALTinterwordspacing
Z.~Ma, C.~Peng, P.~Gao, X.~Meng, Y.~Zou, and B.~Xie, ``Sorft: Issue resolving with subtask-oriented reinforced fine-tuning,'' 2025. [Online]. Available: \url{https://arxiv.org/abs/2502.20127}
\BIBentrySTDinterwordspacing

\bibitem{hui2024qwen25codertechnicalreport}
\BIBentryALTinterwordspacing
B.~Hui, J.~Yang, Z.~Cui, J.~Yang, D.~Liu, L.~Zhang, T.~Liu, J.~Zhang, B.~Yu, K.~Lu, K.~Dang, Y.~Fan, Y.~Zhang, A.~Yang, R.~Men, F.~Huang, B.~Zheng, Y.~Miao, S.~Quan, Y.~Feng, X.~Ren, X.~Ren, J.~Zhou, and J.~Lin, ``Qwen2.5-coder technical report,'' 2024. [Online]. Available: \url{https://arxiv.org/abs/2409.12186}
\BIBentrySTDinterwordspacing

\bibitem{sheng2024hybridflow}
G.~Sheng, C.~Zhang, Z.~Ye, X.~Wu, W.~Zhang, R.~Zhang, Y.~Peng, H.~Lin, and C.~Wu, ``Hybridflow: A flexible and efficient rlhf framework,'' \emph{arXiv preprint arXiv: 2409.19256}, 2024.

\end{thebibliography}

\end{document}